\documentclass[%
 reprint,
 amsmath,amssymb,
 aps,
]{revtex4-2}
\usepackage{xcolor}
\usepackage{graphicx}% Include figure files
\usepackage{dcolumn}% Align table columns on decimal point
\usepackage{bm}% bold math
\usepackage{natbib}
\usepackage{hyperref}

\begin{document}

\preprint{APS/123-QED}

\title{First-Order Perturbations of Covariant Maxwell Equations in Gravitational Waves}% Force line breaks with \\

\author{Lingyue Lou$^{1}$}
\author{Haorong Wu$^{1}$}
\author{Xi-Long Fan$^{1}$}%
 \email{E-mail:xilong.fan@whu.edu.cn}

\affiliation{$^{1}$School of Physics and Technology, Wuhan University, Wuhan 430072, China}
\date{\today}% It is always \today, today,
             %  but any date may be explicitly specified

\begin{abstract}
We present a systematic theoretical framework for investigating first-order electromagnetic (EM) perturbations induced by gravitational waves (GWs). Beginning with the covariant Maxwell equations, we derive the complete first-order perturbation equations in terms of both the EM field tensor and the four-potential, demonstrating their equivalence alongside the residual gauge invariance under the Lorenz gauge condition. Furthermore, explicit first-order expressions for the induced electric and magnetic fields, as well as the associated EM energy-momentum tensor, are obtained.  As an explicit illustration, we analytically evaluate the interaction between a plane EM wave and a GW within the transverse-traceless gauge. By demonstrating that the maximum modulus of the coupling coefficient is on the order of $10^2$, we quantitatively establish that a typical astrophysical GW with a dimensionless strain of $h_0 \sim 10^{-21}$ generates a first-order EM response on the order of $10^{-19}$ relative to the incident field amplitude.
\end{abstract}

%\keywords{Suggested keywords}%Use showkeys class option if keyword
                              %display desired

%\tableofcontents
\maketitle

\section{INTRODUCTION}

The historic direct observation of gravitational waves (GWs) from a binary black hole merger by the LIGO and Virgo collaborations \cite{2016PhRvL.116f1102A} established the foundation for GW astronomy and catalyzed the advent of multi-messenger astrophysics. While current ground-based interferometers operate with extraordinary sensitivity in the frequency band of $10\text{ Hz}$ to $10\text{ kHz}$, and upcoming observatories such as LISA \cite{2006PhRvD..73f4030B,1996CQGra..13A.247D,2003CQGra..20S...1D} and Cosmic Explorer \cite{2021PhRvD.103l2004H,2021arXiv210909882E} are designed to probe lower frequency regimes, the spectrum beyond $10\text{ kHz}$ remains largely unexplored. High-frequency GWs exceeding $10\text{ kHz}$ are of particular physical interest, as they are predicted to arise from exotic cosmological and astrophysical phenomena, including early-universe phase transitions, primordial black hole evaporation, and beyond-Standard-Model physics \cite{2025LRR....28...10A}.

Detecting high-frequency GWs presents a formidable challenge. Alternative detection mechanisms have been extensively proposed, predominantly focusing on the electrodynamic response of EM (EM) fields to gravitational perturbations. The fundamental premise is that GWs traversing a background EM field can induce perturbative electric and magnetic signals \cite{gertsenshteln_wave_nodate}. This inverse Gertsenshtein effect has inspired numerous innovative detector concepts, including microwave cavities \cite{1974ZhETF..65.1729B,1975ZhETF..68.1569G,1979GReGr..11..407B,1975ZhETF..68.1569G,2007arXiv0712.3721M,1979PhLB...80..323C,1978PhLA...68..165P,2003CQGra..20.3505B,2000CQGra..17.2525C,2022PhRvD.105k6011B,2022PhRvD.106j4003Z,2025arXiv251220592S}, coupling EM detecting systems for signal photon flux \cite{2009PhRvD..80f4013L,2018PhRvD..98f4028Z,2026arXiv260107179W,2025PhRvD.112l4002W}, and cutting-edge quantum sensing technologies aiming to detect gravitons \cite{2024NatCo..15.7229T,2025PhRvL.135q1501S,2026PhRvR...8a3140K}. Concurrently, in astrophysical contexts, the interaction between GWs and strong ambient magnetic fields—such as those around neutron stars—can convert gravitational energy into observable radio signals \cite{2024MNRAS.527.4378K,2025ApJ...990..156H, 2025JCAP...11..016D}. 

To accurately predict the outcome of these highly sensitive detection schemes, a rigorous and robust theoretical framework governing the EM-GW interaction is indispensable. The propagation and behavior of EM fields in a gravitational background are fundamentally dictated by the covariant Maxwell equations in curved spacetime \cite{1970NCimB..70..129B,1998CQGra..15.2493M,2017EPJC...77..237C,2005CQGra..22..393T,2023PDU....4001187P,2025EPJC...85..240R}. Various theoretical approaches have been developed to study this interaction, often employing perturbation theory to deduce the response of specific EM configurations \cite{1999CQGra..16..643C,2021EPJC...81..563K,2021EPJC...81...95P,2025ApJ...985..137L}. However, despite these advancements, the analytical treatment of first-order EM perturbations demands meticulous attention to tensorial consistency and gauge invariance. When applying metric perturbations $h_{\mu\nu}$, the operations of raising and lowering tensor indices intrinsically introduce additional coupling terms that break the apparent symmetry between covariant and contravariant field representations. 

In this paper, we employ a systematic perturbation expansion method to thoroughly analyze the behavior of EM (EM) fields within gravitational wave (GW) backgrounds, strictly grounded in a covariant tensor formalism. We rigorously formulate the first-order perturbative field equations and demonstrate their equivalence whether expressed in terms of the field strength tensor or the four-potential. Furthermore, we verify that these first-order equations maintain gauge invariance under the Lorenz gauge. From fundamental definitions, we derive the corresponding first-order expressions for the electric field, magnetic field, and energy-momentum tensor. Finally, by imposing appropriate initial and boundary conditions, the first-order four-potential for a given physical model can be solved using the Green's function method, from which the corresponding electric and magnetic fields can be explicitly calculated.

The structure of this paper is organized as follows. In Section II, we present the perturbative expansion of Maxwell equations in curved spacetime, derive the first-order equations for the four-potential, rigorously demonstrate the invariance of these equations under the Lorenz gauge, and obtain the first-order expressions for the electric field, magnetic field, and energy-momentum tensor. In Section III, we solve these perturbation equations analytically for the interaction between a plane EM wave and a GW in the transverse-traceless (TT) gauge, explicitly discussing the extraction of physical transverse components. Finally, our conclusions and the broader implications of this framework are summarized in Section IV.

Throughout this paper, we adopt the standard tensor index notation and employ geometric units where the speed of light is set to unity ($c=1$). The spacetime metric is chosen to have the signature $(-, +, +, +)$. Greek indices (e.g., $\mu, \nu, \rho, \lambda$) run from 0 to 3, while Latin indices (e.g., $i, j, k, m$) run from 1 to 3 to denote spatial components.

\section{EM Field in Curved Spacetime}

\begin{figure*}[t]
    \centering
    \includegraphics[width=\textwidth]{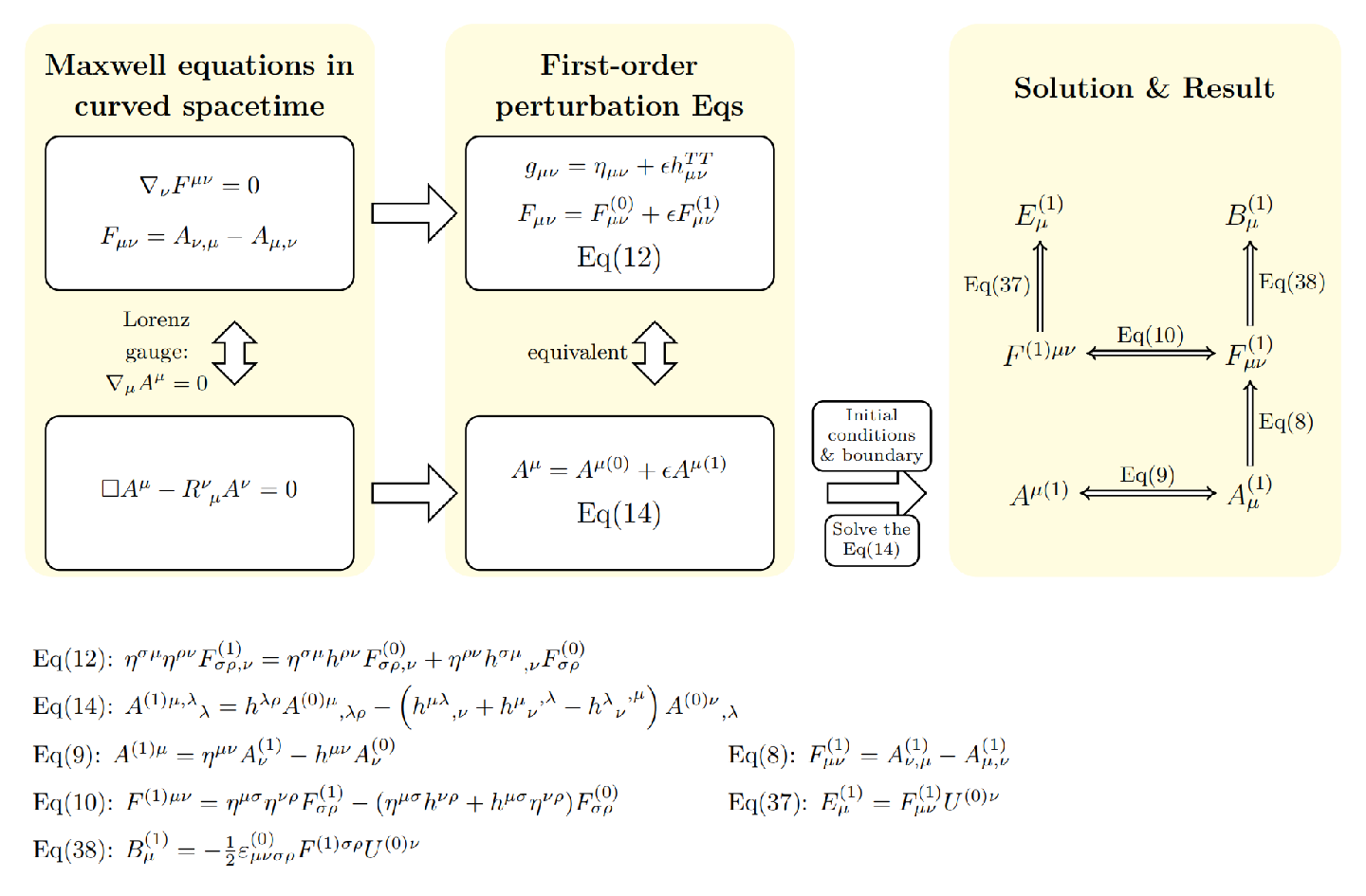}
    \caption{The conceptual framework and computational logic of our calculation. We start from the Maxwell equations in curved spacetime and adopt the Lorenz gauge. We then use a perturbation expansion to obtain the first-order perturbation equations and solve for the first-order vector potential $A^{\mu(1)}$. Finally, we derive the first-order electric and magnetic fields via index raising/lowering and expansion based on their definitions.}
    \label{fig:A1.png}    
\end{figure*}
In this section, we employ the perturbation expansion method to analyze the behavior of EM fields within GW backgrounds, starting from the Maxwell equations in curved spacetime. We formulate the first-order perturbative field equations in a covariant tensor form and elucidate the relationships between the underlying physical quantities. Finally, we derive explicit analytical expressions for the first-order perturbed electric and magnetic fields, as well as the corresponding energy-momentum tensor. The conceptual framework and computational logic of this analysis are summarized in Figure \ref{fig:A1.png}.

\subsection{Prorogation of EM Fields in Curved Spacetime}
In curved spacetime, the prorogation eq of   EM fields is governed by the covariant form of  source-free Maxwell equations, which can be expressed as
\begin{equation} 
    \nabla_\nu F^{\mu\nu} = 0\,,\label{ME}
\end{equation} 
where $F^{\mu\nu}$ is the contravariant EM field tensor and $\nabla_\nu$ denotes the covariant derivative. This equation can also be written in terms of partial derivatives of the convariant strength tensor $F_{\mu\nu}$ as
\begin{equation} 
    \frac{1}{\sqrt{-g}} \partial_{\nu}\left(\sqrt{-g} g^{\sigma \mu} g^{\rho \nu} F_{\sigma \rho}\right)=0\,,
\end{equation} 
where $g = \det(g_{\mu\nu})$. Additionally, the EM field tensor satisfies the Bianchi identity
\begin{equation} 
    F_{[\mu \nu, \alpha]} = 0\,.
\end{equation}

The convariant EM field tensor $F_{\mu\nu}$ is related to the  convariant  the four-potential $A_{\mu}$ :  $F_{\mu\nu}= A_{\nu,\mu}- A_{\mu,\nu}$. By introducing   he Lorenz gauge $\nabla_{\mu}A^{\mu}=0$, the Maxwell equations, could also write in term of four-potential:
\begin{equation} \label{eq_wave_A}
        \Box_g A^{\mu} - {R^{\mu}}_{\nu} A^{\nu} = 0\,,
\end{equation}
where $\Box_g \equiv g^{\nu \rho} \nabla_{\nu} \nabla_{\rho}$ denotes the d'Alembertian operator, and $R_{\mu\nu}$ is the Ricci tensor.

\subsection{The perturbation expansion of fields} 

Considering  perturbations caused by GWs, we can expand the metric as
\begin{equation}
    g_{\mu \nu}= \eta_{\mu\nu} + \epsilon h^{TT}_{\mu\nu} \quad |h_{\mu\nu}| \ll 1\,,\label{metric per}
\end{equation}
where $\epsilon$ is a small parameter that labels the order of perturbation and $h^{TT}_{\mu\nu}$ denotes the GW perturbation, which satisfies the transverse-traceless (TT) gauge 
\begin{equation}
    h_{0\mu}={h^\mu}_{\mu}=\partial^\mu h_{\mu\nu}=0,\label{gaugeh}\,.
\end{equation}

To systematically investigate the EM perturbations, we first introduce the four-potential $A^{\mu}$ and the EM field tensor $ F_{\mu\nu}$, and decompose them into a flat spacetime background term and the GW-induced first-order perturbation term:
\begin{equation}
  \begin{aligned}
     A^{\mu} &= A^{\mu(0)} + \epsilon A^{\mu(1)}\,,\quad & \left| A^{\mu(1)} \right| \ll  \left| A^{\mu(0)} \right|\,\\ 
     F_{\mu\nu} &= F_{\mu\nu}^{(0)} + \epsilon F_{\mu\nu}^{(1)}\,, \quad & \left|F_{\mu\nu}^{(1)} \right| \ll \left| F_{\mu\nu}^{(0)} \right| 
      \end{aligned}
\end{equation}
 This directly leads to the following perturbation decomposition for the EM field tensor    
\begin{equation}
    \begin{aligned}
        F_{\mu\nu}&= A_{\nu,\mu}- A_{\mu,\nu}\,,\\
        0: \quad F^{(0)}_{\mu\nu}&=A^{(0)}_{\nu,\mu}- A^{(0)}_{\mu,\nu}\,, \\
        \epsilon: \quad  F^{(1)}_{\mu\nu}&= A^{(1)}_{\nu,\mu}-A^{(1)}_{\mu,\nu}\label{F-A}\,.
    \end{aligned}
\end{equation}

In curved spacetime, index manipulation is performed using the metric tensor $g_{\mu\nu}$ and its inverse $g^{\mu\nu}$. For the contravariant four-potential $A^\mu$, this gives
\begin{equation}
    \begin{aligned}
        A^{\mu}&=g^{\mu\nu}A_{\nu}\,,\\
    0:\quad A^{(0)\mu}&=\eta^{\mu\nu}A^{(0)}_{\nu}\,,\\
    \epsilon:\quad A^{(1)\mu}&=\eta^{\mu\nu}A^{(1)}_\nu-h^{\mu\nu}A^{(0)}_{\nu}\label{RLA}\,.
    \end{aligned}
\end{equation}
Meanwhile,  contravariant form of the EM field tensor and its perturbation expansion are given by
\begin{equation}
    \begin{aligned}
        F^{\mu\nu}&=g^{\mu\sigma}g^{\nu\rho}F_{\sigma\rho}\\
        0: \quad F^{(0)\mu\nu}&=\eta^{\mu\sigma}\eta^{\nu\rho}F^{(0)}_{\sigma\rho}\,, \\
        \epsilon: \quad  F^{(1)\mu\nu}&=\eta^{\mu\sigma}\eta^{\nu\rho}F^{(1)}_{\sigma\rho}- (\eta^{\mu\sigma}h^{\nu\rho}+h^{\mu\sigma}\eta^{\nu\rho})F^{(0)}_{\sigma\rho}\label{Fsj}\,.
    \end{aligned}
\end{equation}

It is important to note that the relation between the four-potential and the EM field tensor is fundamentally defined with covariant form (Eq.~ \ref{F-A}). Moreover, raising or lowering indices of first-order perturbative quantities generally introduces additional terms induced by the coupling to the metric perturbation (Eq.~\ref{RLA} and Eq.~\ref{Fsj}). 
Especially, the $F^{(1)\mu\nu}$ and  $F^{(1)}_{\sigma\rho}$ do not have the same form, which is critical for perturbation expansion of  the electric field $E^{(1)}_\mu$ and the magnetic field $B^{(1)}_\mu$ (see Sec.~\ref{sec_b_e})

\subsection{The perturbation equation}
Under the perturbation expansions of $ g_{\mu \nu}$, $F_{\mu\nu}$ and $A^{\mu}$, Maxwell equations (Eq. \ref{ME}) can be expanded perturbatively. The zeroth-order term reproduces the standard flat spacetime Maxwell equations,   
\begin{equation}
    0: \quad \eta^{\sigma \mu} \eta^{\rho \nu}  {F_{\sigma \rho,\nu}^{(0)}} = 0\,,\label{eqF0}
\end{equation}
 while the first-order term yields the perturbation equation driven by the metric variation:
 \begin{equation}
    \epsilon: \quad \eta^{\sigma \mu} \eta^{\rho \nu}  F_{\sigma \rho,\nu}^{(1)} = \eta^{\sigma\mu} h^{\rho\nu}  F_{\sigma\rho,\nu}^{(0)} + \eta^{\rho\nu}  {h^{\sigma\mu}}_{,\nu} F_{\sigma\rho}^{(0)}\,.\label{Fper1} 
 \end{equation}

 Expanding the  Maxwell equations (Eq. \ref{eq_wave_A}) up to linear order in $h_{\mu\nu}$, we obtain the unperturbed zeroth-order equation  in term of $A^{\mu}$
\begin{equation}
        0: \quad A^{(0)\mu, \lambda}{ }_{\lambda}=0\,,\label{eqA0}
\end{equation}
and the corresponding first-order perturbation equation
\begin{equation}
    \begin{aligned}
        \epsilon: \quad {A^{(1)\mu, \lambda}}_{\lambda}= & h^{\lambda \rho} {A^{(0)\mu}}_{,\lambda \rho}-\\ 
       & \left({h^{\mu \lambda}}_{,\nu}+ {{h^\mu}_\nu}^{,\lambda}-{{h^{\lambda}}_{\nu}}^{,\mu}    \right) {A^{(0)\nu}}_{,\lambda}\,.\label{Aper1}
    \end{aligned}
\end{equation}

Evidently, the perturbation equation (\ref{Fper1}) is equivalent to Eq. (\ref{Aper1}).  This equivalence can be established by utilizing Eq. (\ref{F-A}) and handling the raising and lowering of indices in the perturbation context. 
Substituting Eq. (\ref{F-A}) into Eq. (\ref{Fper1}), we expand both sides
\begin{align} 
    \eta^{\sigma \mu} \eta^{\rho \nu}  F_{\sigma \rho,\nu}^{(1)} &= \eta^{\sigma \mu} \eta^{\rho \nu} \partial_{\nu}(A^{(1)}_{\rho,\sigma} - A^{(1)}_{\sigma,\rho})\,,\label{F-A1}\\ 
    \eta^{\sigma\mu}h^{\rho\nu}F^{(0)}_{\sigma\rho,\nu} &= {h^{\nu}}_{\rho}({A^{(0)\rho,\mu}}_{\nu} - {A^{(0)\mu,\rho}}_{\nu})\,,\label{F-A2}\\ 
    \eta^{\rho\nu}{h^{\sigma\mu}}_{,\nu}F^{(0)}_{\sigma\rho} &= {h^{\mu}}_{\sigma,\nu}(A^{(0)\nu,\sigma} - A^{(0)\sigma,\nu})\,.\label{F-A3} 
\end{align}
Applying the index relations from Eq. (\ref{RLA}) to the right-hand side of Eq. (\ref{F-A1}), we expand the first term 
\begin{equation} 
    \begin{aligned} 
        &\eta^{\sigma \mu}\eta^{\rho\nu}\partial_\nu\partial_\sigma A^{(1)}_\rho\\ 
        =&\eta^{\sigma \mu}\partial_\nu\partial_\sigma [\eta^{\rho\nu}A^{(1)}_\rho]\\ 
        =&\eta^{\sigma \mu}\partial_\nu\partial_\sigma [A^{(1)\nu}+h^{\rho\nu}A^{(0)}_{\rho}]\\ 
        =&{A^{(1)\nu,\mu}}_{\nu}+{h^{\rho\nu,\mu}}_{\nu}A^{(0)}_{\rho}+h^{\rho\nu,\mu}A^{(0)}_{\rho,\nu}\\ 
        &+{h^{\rho\nu}}_{,\nu}A^{(0),\mu}_{\rho}+h^{\rho\nu}{{A_{\rho}^{(0),\mu}}}_{\nu}\,, 
    \end{aligned} 
\end{equation} 
and the second term 
\begin{equation} 
    \begin{aligned} 
        &\eta^{\sigma\mu}\eta^{\rho\nu}\partial_{\nu}\partial_{\rho}A^{(1)}_\sigma\\ 
        =&\eta^{\rho\nu}\partial_{\nu}\partial_{\rho}[\eta^{\sigma\mu}A^{(1)}_{\sigma}]\\ 
        =&\eta^{\rho\nu}\partial_{\nu}\partial_{\rho}[A^{(1)\mu}+h^{\sigma\mu}A^{(0)}_{\sigma}]\\ 
        =&{A^{(1)\mu,\nu}}_{\nu}+{h^{\sigma\mu,\nu}}_{\nu}A^{(0)}_{\sigma}+2h^{\sigma\mu,\nu}A^{(0)}_{\sigma,\nu}\\ 
        &+h^{\sigma\mu}{A_{\sigma}^{(0),\nu}}_{\nu}\,. 
    \end{aligned} 
\end{equation} 
Substituting these expansions back into Eq.~(\ref{F-A1}) and simplifying using the TT gauge conditions for GWs (Eq.~\ref{gaugeh}) and the Lorenz gauge condition for the EM field, we obtain:
\begin{equation} 
    \begin{aligned} 
        \eta^{\sigma \mu} \eta^{\rho \nu} \partial_{\nu} F_{\sigma \rho}^{(1)} =& -{A^{(1)\mu,\nu}}_{\nu} - 2{{h^{\mu}}_{\sigma}}^{,\nu}{A^{(0)\sigma}}_{,\nu}\\ 
        &+ {{h^{\nu}}_{\rho}}^{,\mu}{A^{(0)\rho}}_{,\nu} + {h^{\nu}}_{\rho}{A^{(0)\rho,\mu}}_{\nu}\,. \label{F1-A1}
    \end{aligned} 
\end{equation}

Now the three terms of  first-order perturbation equation   Eq.~(\ref{Fper1}) are expressed in terms of $A^{\mu}$. Substituting the expanded forms from Eqs.~(\ref{F1-A1}), (\ref{F-A2}), and (\ref{F-A3}) into Eq.~(\ref{Fper1}), one can get the 
first-order perturbation equation  Eq.~(\ref{Aper1}).

\subsection{Gauge Invariance of the First-Order Perturbation Equations}
\label{gaugesection}

In this subsection, we demonstrate that the first-order perturbed equation (\ref{Aper1}) are invariant under residual gauge transformations within the Lorenz gauge. The four-dimensional vector potential admits a gauge transformation of the form $\tilde{A}_{\mu} = A_{\mu} + \nabla_\mu \chi$. Expanding this perturbatively yields:
\begin{equation}
    \begin{aligned}
        0:\quad \tilde{A}^{(0)}_{\mu} &= A^{(0)}_{\mu} + \partial_{\mu}\chi^{(0)}\,,\\
        \epsilon:\quad \tilde{A}^{(1)}_{\mu} &= A^{(1)}_{\mu} + \partial_{\mu}\chi^{(1)}\,.\label{gauge}
    \end{aligned}
\end{equation}
Under this transformation, it is straightforward to verify that the first-order EM field tensor $F^{(1)}_{\mu\nu}$ remains invariant
\begin{equation}
    \begin{aligned}
    \tilde{F}^{(1)}_{\mu\nu} &= \partial_{\mu}\tilde{A}^{(1)}_{\nu} - \partial_{\nu}\tilde{A}^{(1)}_{\mu}\\
    &= \partial_{\mu}(A^{(1)}_{\nu} + \partial_{\nu}\chi^{(1)}) - \partial_{\nu}(A^{(1)}_{\mu} + \partial_{\mu}\chi^{(1)})\\
    &= \partial_{\mu}A^{(1)}_{\nu} - \partial_{\nu}A^{(1)}_{\mu} = F^{(1)}_{\mu\nu}\,.
    \end{aligned}
\end{equation}
Consequently, by applying the gauge transformation (\ref{gauge}) and the index-raising/lowering rules (\ref{RLA}), the transformed contravariant vector potential to first order is given by
\begin{equation}
    \begin{aligned}
        0:\quad \tilde{A}^{(0)\mu} &= A^{(0)\mu} + \partial^{\mu}\chi^{(0)}\,,\\
        \epsilon:\quad \tilde{A}^{(1)\mu} &= \eta^{\mu\nu}A^{(1)}_{\nu} - h^{\mu\nu}A^{(0)}_{\nu} + \partial^{\mu}\chi^{(1)} - h^{\mu\nu}\partial_{\nu}\chi^{(0)}\\
        &= A^{(1)\mu} + \partial^{\mu}\chi^{(1)} - h^{\mu\nu}\partial_{\nu}\chi^{(0)}\,.
    \end{aligned}
\end{equation}

We now consider the variation of the first-order perturbed equation, Eq.~(\ref{Aper1}), under this gauge transformation. The governing equation is
\begin{equation}
    \begin{aligned}
        \epsilon: \quad {\tilde{A}^{(1)\mu, \lambda}}_{\qquad\,\lambda}= & h^{\lambda \rho} {\tilde{A}^{(0)\mu}}_{\quad\,\,,\lambda \rho}-\\ 
       & \left({h^{\mu \lambda}}_{,\nu}+ {{h^\mu}_\nu}^{,\lambda}-{{h^{\lambda}}_{\nu}}^{,\mu}    \right) {\tilde{A}^{(0)\nu}}_{\quad\,\,,\lambda}\,.
    \end{aligned}
\end{equation}
Substituting the transformation into the left-hand side (LHS) of the variation gives
\begin{equation} \label{left}
    \begin{aligned}
        \Delta \text{LHS} &= \partial_{\lambda}\partial^{\lambda}[\partial^{\mu}\chi^{(1)} - h^{\mu\nu}\partial_{\nu}\chi^{(0)}]\\
        &= \partial_{\lambda}\partial^{\lambda}\partial^{\mu}\chi^{(1)} - 2\partial_{\lambda}h^{\mu\nu}\partial^{\lambda}\partial_{\nu}\chi^{(0)}\,.
    \end{aligned}
\end{equation}
Note that in obtaining Eq.~(\ref{left}), we applied the properties of the TT gauge where the GW equation satisfies $\Box h^{\mu\nu} = 0$, along with the zeroth-order gauge condition $\Box \chi^{(0)} = \partial_{\lambda}\partial^{\lambda}\chi^{(0)} = 0$, to eliminate the remaining terms.

Similarly, the variation of the right-hand side (RHS) is evaluated as
\begin{equation} \label{right}
    \begin{aligned}
        \Delta \text{RHS} =& h^{\lambda\rho}\partial_{\lambda}\partial_{\rho}\partial^{\mu}\chi^{(0)} - ({h^{\mu\lambda}}_{,\nu} + {{h^{\mu}}_{\nu}}^{,\lambda} - {{h^{\lambda}}_{\nu}}^{,\mu})\partial_{\lambda}\partial^{\nu}\chi^{(0)}\\
        =& \partial^{\mu}[h^{\lambda\nu}\partial_{\lambda}\partial_{\nu}\chi^{(0)}] - {{h^{\lambda}}_{\nu}}^{,\mu}\partial_{\lambda}\partial^{\nu}\chi^{(0)}\\
        & - ({h^{\mu\lambda}}_{,\nu} + {{h^{\mu}}_{\nu}}^{,\lambda} - {{h^{\lambda}}_{\nu}}^{,\mu})\partial_{\lambda}\partial^{\nu}\chi^{(0)}\\
        =& \partial^{\mu}(h^{\lambda\nu}\partial_{\lambda}\partial_{\nu}\chi^{(0)}) - 2{h^{\mu\lambda}}_{,\nu}\partial_{\lambda}\partial^{\nu}\chi^{(0)}\,.
    \end{aligned}
\end{equation}
To relate these variations, we must examine the Lorenz gauge condition dictated by the perturbation equations
\begin{equation}
    \nabla_{\mu}\tilde{A}^{\mu} = \nabla_{\mu}A^{\mu}\,.
\end{equation}
This condition restricts the infinitesimal gauge transformation function $\chi$ to satisfy
\begin{equation}
    \Box_g \chi = g^{\mu\nu}(\partial_{\mu}\partial_{\nu}\chi - \Gamma^{\rho}_{\mu\nu}\partial_{\rho}\chi) = 0\,.
\end{equation}
Performing a perturbative expansion of this constraint and noting that the background Christoffel symbols vanish in flat spacetime ($\Gamma^{(0)\rho}_{\mu\nu}=0$), we find
\begin{equation}
    \begin{aligned}
        &g^{\mu\nu}(\partial_{\mu}\partial_{\nu}\chi - \Gamma^{\rho}_{\mu\nu}\partial_{\rho}\chi)\,\\
        0:\quad & \eta^{\mu\nu}\partial_{\mu}\partial_{\nu}\chi^{(0)} = 0\,,\\
        \epsilon:\quad & \eta^{\mu\nu}\partial_{\mu}\partial_{\nu}\chi^{(1)} - \eta^{\mu\nu}\Gamma^{(1)\rho}_{\mu\nu}\partial_{\rho}\chi^{(0)} - h^{\mu\nu}\partial_{\mu}\partial_{\nu}\chi^{(0)} = 0\,.
    \end{aligned}
\end{equation}
In the TT gauge, the contraction of the Minkowski metric with the first-order Christoffel symbol vanishes identically
\begin{equation}
    \eta^{\mu\nu}\Gamma^{(1)\rho}_{\mu\nu} = \frac{1}{2}\eta^{\mu\nu}({h^{\rho}}_{\mu,\nu} + {h^{\rho}}_{\nu,\mu} - {h_{\mu\nu}}^{,\rho}) = 0\,.
\end{equation}
Thus, the constraint on the first-order infinitesimal transformation under the Lorenz gauge simplifies to
\begin{equation}
    \eta^{\mu\nu}\partial_{\mu}\partial_{\nu}\chi^{(1)} = h^{\mu\nu}\partial_{\mu}\partial_{\nu}\chi^{(0)}\,.
\end{equation}
Applying this simplified constraint to the first term of the LHS variation in Eq.~(\ref{left}), we obtain:
\begin{equation}
    \partial_{\lambda}\partial^{\lambda}\partial^{\mu}\chi^{(1)} = \partial^{\mu}(h^{\lambda \nu}\partial_{\lambda}\partial_{\nu}\chi^{(0)})\,.
\end{equation}
This perfectly matches the first term of the RHS variation in Eq.~(\ref{right}). Furthermore, by relabeling the dummy indices ($\lambda \leftrightarrow \nu$), it is straightforward to observe that the second terms also exactly cancel each other out. 

Therefore, we conclude that $\Delta \text{LHS} = \Delta \text{RHS}$, rigorously proving that the first-order perturbation equation is invariant under the residual Lorenz gauge transformation.

\subsection{Representation of Electric and Magnetic Fields in Curved Spacetime}\label{sec_b_e}
For an observer at point $p$ in spacetime with four-velocity $U^{\mu}$ satisfying $g_{\mu\nu}U^{\mu}U^{\nu}=-1$, the measured electric field $E_{\mu}$ and magnetic field $B_{\mu}$ are defined as
\begin{equation} 
    E_\mu=F_{\mu\nu}U^{\nu}\quad B_{\mu}=-\frac{1}{2}\varepsilon_{\mu\nu\sigma\rho} F^{\sigma\rho}U^{\nu}\,, 
\end{equation} 
where $\varepsilon^{}_{\mu\nu\sigma\rho}$ is the four-dimensional Levi-Civita tensor with $\varepsilon^{}_{\mu\nu\sigma\rho}=\sqrt{-g}[\mu\nu\sigma\rho]$, and its first-order perturbation term $\varepsilon^{(1)}_{\mu\nu\sigma\rho}=0$. The observer's four-velocity is determined by the geodesic equation, whose perturbation expansion is given by
\begin{equation} 
\begin{aligned} 
        & \frac{dU^\mu}{d\tau} + \Gamma^\mu_{\nu\rho} U^\nu U^\rho  = 0\\ 
         0:\quad & \frac{dU^{(0)\mu}}{d\tau} + \Gamma^{(0)\mu}_{\nu\rho} U^{(0)\nu} U^{(0)\rho}  = 0\,,\\ 
         \epsilon:\quad & \frac{dU^{(1)\mu}}{d\tau}+2\Gamma^{(0)\mu}_{\nu\rho} U^{(0)\nu} U^{(1)\rho}+\Gamma^{(1)\mu}_{\nu\rho}U^{(0)\nu}U^{(0)\rho}  =0\,,
    \end{aligned} 
\end{equation} 
where the perturbation expansion of the Christoffel symbols is
\begin{equation} 
    \begin{aligned} 
        & \Gamma^{\mu}_{\nu \rho}=\frac{1}{2}g^{\mu\sigma}(g_{\sigma\nu,\rho}+g_{\sigma\rho, \nu}-g_{\nu \rho,\sigma})\\ 
        0:\quad & \Gamma^{(0)\mu}_{\nu\rho}=0\,,\\ 
        \epsilon:  \quad & \Gamma^{(1)\mu}_{\nu\rho}=\frac{1}{2}\eta^{\mu\sigma}(h_{\sigma\nu,\rho}+h_{\sigma\rho,\nu}-h_{\nu\rho,\sigma})\,. 
    \end{aligned} 
\end{equation} 
For a stationary observer with $U^{(0)\mu}=(1,0,0,0)$,  we find $dU^{(1)\mu}/d\tau=0$. This implies that $U^{(1)\mu}$ is constant along the observer's worldline. Using the initial condition $ U^{(1)\mu}(\tau \to -\infty) = 0$, which corresponds to the state before the GW arrives, we therefore conclude that $U^{(1)\mu}=0$ for all times. In fact, for GWs in the TT gauge, explicit calculations reveal that all higher-order perturbations of the four-velocity identically vanish. Thus, the observer's four-velocity remains completely unperturbed throughout the entire interaction with the gravitational wave.
Additionally, the covariant form of the four-velocity and its perturbation expansion are
\begin{equation}
    \begin{aligned}
        U_{\mu} & =U^{\nu}g_{\mu\nu}\\
        0:\quad U^{(0)}_\mu & =U^{(0)\nu}\eta_{\mu\nu}\,,\\
        \epsilon:\quad U^{(1)}_\mu &=U^{(0)\nu}h_{\mu\nu}+U^{(1)\nu}\eta_{\mu\nu}\,.
    \end{aligned}
\end{equation}
Since $U^{(1)\nu}=0$, we can similarly conclude that $U^{(1)}_\mu=0$. We can now expand the electric field perturbatively, which simplifies to
\begin{equation} 
    \begin{aligned}
        E_\mu&=F_{\mu\nu}U^{\nu}\\
        0: \quad E^{(0)}_{\mu}&=F^{(0)}_{\mu\nu}U^{(0)\nu}\,, \\ 
        \epsilon: \quad  E^{(1)}_\mu&=F^{(1)}_{\mu\nu}U^{(0)\nu}\label{E1}\,. 
    \end{aligned}
\end{equation}
Similarly, the magnetic field perturbation expansion (using $\varepsilon^{(1)}_{\mu\nu\sigma\rho}=0$) becomes
\begin{equation}
    \begin{aligned}
         B_{\mu}&=-\frac{1}{2}\varepsilon_{\mu\nu\sigma\rho} F^{\sigma\rho}U^{\nu}\\
        0: \quad B^{(0)}_{\mu}&=-\frac{1}{2}\varepsilon^{(0)}_{\mu\nu\sigma\rho} F^{(0)\sigma\rho}U^{(0)\nu}\,, \\ 
        \epsilon: \quad  B^{(1)}_\mu&=-\frac{1}{2}\varepsilon^{(0)}_{\mu\nu\sigma\rho} F^{(1)\sigma\rho}U^{(0)\nu}\label{B1}\,. 
    \end{aligned}
\end{equation}
where $F^{(1)\sigma\rho}$ is given by Eq. (\ref{Fsj}). It should be emphasized that, precisely because the covariant and contravariant forms of the first-order EM field tensor do not follow the simple index raising and lowering relations analogous to those in flat spacetime (see Eq. \ref{Fsj}), the first-order magnetic field consequently cannot be expressed in a simple matrix expansion form. Consequently, the wave equations governing $B^{(0)}_{\mu}$ and $B^{(1)}_\mu$ cannot be identical in form.

\subsection{Energy-Momentum Tensor of the EM Field in Curved Spacetime}

The energy-momentum tensor for the EM field in curved spacetime is given by
\begin{equation} 
    T^{\mu \nu} = \frac{1}{4\pi} \left[g_{\rho\sigma}F^{\mu \rho}F^{\nu\sigma} - \frac{1}{4} g^{\mu \nu} F_{\rho\sigma} F^{\rho\sigma} \right] \,,
\end{equation}
where all tensor components are treated as real quantities. We expand $T^{\mu\nu}$ perturbatively as $T^{\mu\nu} = T^{(0)\mu\nu} +\epsilon T^{(1)\mu\nu} + \epsilon^2\ T^{(2)\mu\nu} + \dots$. The zeroth-order term, representing the energy-momentum tensor in flat spacetime, is
\begin{equation} 
    T^{(0)\mu \nu} =\frac{1}{4\pi}[\eta_{\rho\sigma}F^{(0)\mu\rho}F^{(0)\nu\sigma}-\frac{1}{4}\eta^{\mu\nu}F^{(0)}_{\rho\sigma}F^{(0)\rho\sigma}]\,,
\end{equation} 
the first-order perturbation term is derived as
\begin{equation} 
    \begin{aligned} 
        T^{(1)\mu\nu} =&\frac{1}{4\pi}[h_{\rho\sigma}F^{(0)\mu\rho}F^{(0)\nu\sigma}+2\eta_{\rho\sigma}F^{(1)\mu\rho}F^{(0)\nu\sigma}\\ 
        &+\frac{1}{4}h^{\mu\nu}F^{(0)}_{\rho\sigma}F^{(0)\rho\sigma}-\frac{1}{2}\eta^{\mu \nu}F^{(1)}_{\rho\sigma}F^{(0)\rho\sigma}]\,,\label{T1}
    \end{aligned} 
\end{equation}
and the second-order perturbation term 
\begin{equation} 
    \begin{aligned} 
        T^{(2)\mu\nu} &=\frac{1}{4\pi}[\eta_{\rho \sigma}(F^{(1)\mu\rho}F^{(1)\nu\sigma}+2F^{(0)\mu\rho}F^{(2)\nu\sigma})\\
        &+2h_{\rho\sigma}F^{(0)\mu\rho}F^{(1)\nu\sigma}-\frac{1}{4}\eta^{\mu\nu}(F^{(1)}_{\rho \sigma}F^{(1)\rho\sigma}\\
        &+2F^{(0)}_{\rho\sigma}F^{(2)\rho\sigma})+\frac{1}{2}h^{\mu\nu}F^{(0)}_{\rho\sigma}F^{(1)\rho\sigma}]\,.\label{T2}
    \end{aligned} 
\end{equation}
The energy density, defined as $\rho = T^{\mu\nu}U_{\mu}U_{\nu}$, and the Poynting vector, which represents the energy flux of the EM field, is defined as
\begin{equation}
    S^{\mu} = -T^{\mu\nu}U_{\nu}-\rho U^{\mu}\,.
\end{equation}
Since the perturbations of the four-velocity are all zero, we can readily obtain that the corresponding perturbation of the energy density is
\begin{equation}
    \begin{aligned}
        \epsilon:\quad \rho^{(1)} &=T^{(1)tt}\,,\\
        \epsilon^2:\quad \rho^{(2)} &=T^{(2)tt}\,,
    \end{aligned}
\end{equation}
and the perturbation of the spatial components of the energy flux is
\begin{equation}
    \begin{aligned}
        \epsilon:\quad S^{(1)i} &=T^{(1)it}\,,\\
        \epsilon^2:\quad S^{(2)i} &=T^{(2)it}\,,
    \end{aligned}
\end{equation}
where the index $i$ denotes the spatial components. Similarly, the calculation of the energy flux also differs from the expression in flat spacetime; for example, the first-order energy flux expressed in terms of the electric and magnetic fields is
\begin{equation}
    \begin{aligned}
    S^{(1)i}=&\frac{1}{4\pi}\varepsilon^{(0)}_{mjk}[-h^{im}E^{(0)j}B^{(0)k}\\
    &+\eta^{im}E^{(1)j}B^{(0)k}+\eta^{im}E^{(0)j}B^{(1)k}]\,.
    \end{aligned}
\end{equation}

\section{The solution of the  perturbation expansion }\label{sec_solve_eq}
To evaluate the first-order EM perturbations induced by GWs, several analytical frameworks can be employed. One common method directly formulates wave equations for the perturbed electric and magnetic fields (e.g., \cite{2003PhRvD..67j4008L, 2022PhRvD.106j4003Z}). Another approach utilizes Maxwell's equations in terms of the EM field tensor to derive the corresponding perturbation equations of electric and magnetic  fields (e.g.\cite{2025arXiv250421225A}). However, these treatments often assume that the wave equations governing the background fields $E^{(0)}_{\mu }$ ($B^{(0)}_{\mu}$) and the first-order fields $E^{(1)}_\mu$ ($B^{(1)}_\mu$) share an identical functional form, an assumption that contradicts our derived Eqs.~(\ref{E1}), (\ref{B1}), and (\ref{Fsj}). Because the explicit differential equations for $E^{(1)}_\mu$ and $B^{(1)}_\mu$ are highly intricate and resist compact closed-form representation, solving them directly is mathematically impractical. While Eq.~(\ref{Fper1}) provides a valid first-order system for the field strength tensor $F_{\sigma \rho}$, reformulating the problem in terms of the four-potential $A^{\mu}$ via Eq.~(\ref{Aper1}) is significantly more computationally tractable. This approach naturally reduces the system to a set of decoupled inhomogeneous wave equations for each vector component. Consequently, we adopt Eq.~(\ref{Aper1}) as the fundamental basis for our subsequent analytical treatment.

By solving the perturbation equation Eq.~(\ref{Aper1}), one can obtain $A^{(1)\mu}$. Using the index-lowering relation (\ref{RLA}) together with the definition (\ref{F-A}), the corresponding EM field tensor $F^{(1)}_{\mu\nu}$ can then be derived. Subsequently, the first-order electric and magnetic fields are obtained from the definitions (\ref{E1}) and (\ref{B1}). Furthermore, the corresponding first- and second-order energy-momentum tensors can be calculated from Eqs.~(\ref{T1}) and (\ref{T2}). A crucial point to emphasize is that when calculating the first-order perturbation of the EM field in the presence of GWs, the selected EM field must first satisfy the zeroth-order conditions of Maxwell equations (either Eq. \ref{eqF0} or Eq. \ref{eqA0}). Only then can it be substituted into the first-order perturbation equations for calculation.

The first-order perturbation equation could be written as
\begin{equation}\label{eqinhomA}
    \Box_{\eta} A^{(1)\mu} = J^\mu_{\text{eff}}\,,
\end{equation}
where $J^\mu_{\text{eff}}$ refers to the effective source on the right-hand side of Eq.~(\ref{Aper1}). This inhomogeneous equation can be formally solved using the Green's function method, yielding the integral solution

\begin{equation}
    A^{(1)\mu}(\vec r,t)=\int dt' \int_V d^3\vec r' \,G(\vec r,t;\vec r',t')J^{\mu}_{\text{eff}}(\vec r',t')\,,
\end{equation}
where the Green's function $G(\vec r,t;\vec r',t')$ satisfies the defining differential equation
\begin{equation}
    \Box_{\eta} G(\vec r,t;\vec r',t') = \delta(t-t')\delta^{(3)}(\vec r-\vec r')\,.
\end{equation}

\subsection{Case Study}
To demonstrate our approach,  we consider a GW in the TT gauge, propagating in the $y$-$z$ plane at an angle $\theta$ with respect to the positive $z$-axis. The metric perturbation tensor is given by
\begin{align} 
     h_{\mu\nu} = \begin{pmatrix}        0 & 0 & 0 & 0 \\        0 & h_{+} & h_{\times} \cos \theta & -h_{\times} \sin \theta \\        0 & h_{\times} \cos \theta & -h_{+} \cos^2 \theta & h_{+} \sin \theta \cos \theta \\        0 & -h_{\times} \sin \theta & h_{+} \sin \theta \cos \theta & -h_{+} \sin^2 \theta    \label{h1}
      \end{pmatrix} 
\end{align} 
where the polarization states are
\begin{equation}
    \begin{aligned}
         h_+(t,y,z) &= h_{0+} \exp\{i[k_g(y\sin\theta+z\cos\theta)-\omega_g t]\} \,,\\    h_\times(t,y,z) &= h_{0\times} \exp\{i[k_g(y\sin\theta+z\cos\theta)-\omega_g t]\}\,.\label{h2}
    \end{aligned}
\end{equation}
where $h_{0+}$ and $h_{0\times}$ are the amplitudes of the two states of the GW, $k_g$ is the wave vector, and $\omega_g$ is its angular frequency.
We also consider a free plane EMW propagating along the positive $z$-axis with linear polarization along the $y$-axis
\begin{equation} 
    \textbf{A}^{(0)}(t,\textbf{r}) = \frac{E_{0y}}{\omega_a}e^{i(k_az-\omega_at)} \mathbf{e}_y \,,\quad E_{0y}= \text{const}\,, \label{A0}
\end{equation} 
where $E_{0y}$ represents the amplitude of the electric field in the $y$-direction, $k_a$ is the wave vector of the EMW, and $\omega_a$ is its angular frequency.

Although both the EMW and the GW are conveniently expressed in complex form, the physically observable fields are given by their real parts. Substituting the complex representations of the EMW and GW into the first-order perturbation equation (\ref{Aper1}), one finds that the nonlinear interaction term contains products of oscillatory factors of the form $e^{i(\textbf{k}_a \cdot \textbf{r}- \omega_a t)}$ and $e^{i(\textbf{k}_g \cdot \textbf{r}- \omega_g t)}$. In analogy with wave-mixing processes, these products naturally decompose into two distinct frequency components through the identity for the product of harmonic waves. Consequently, the source term excites both a sum-frequency mode and a difference-frequency mode, corresponding respectively to frequencies
\begin{equation}
    \begin{aligned}
        \omega_{+} & = \omega_a + \omega_g\,,\quad \textbf{k}_+=\textbf{k}_a+\textbf{k}_g\,;\\
        \omega_{-} & = \omega_a - \omega_g\,,\quad \textbf{k}_-=\textbf{k}_a-\textbf{k}_g\,.
    \end{aligned}
\end{equation}

Here we present only the calculation results for the sum-frequency mode below.
Substituting the assumed conditions (Eq. \ref{h1} and Eq. \ref{A0}) into the perturbation equation (\ref{Aper1}) and expanding by indices, we obtain
\begin{subequations}\label{EQ}
\begin{align}
    &\Box_{\eta} A^{(1)t} = -E_{0y}h_{0+}\omega_g\cos\theta\sin\theta e^{i\Phi}\,, \\   
    &\Box_{\eta} A^{(1)x} =-E_{0y}h_{0\times}\omega_g(\cos{\theta}-\cos{2\theta})e^{i\Phi}\,, \\   
    &\Box_{\eta} A^{(1)y} =E_{0y}h_{0+}(\omega_a\sin^2\theta+\omega_g\cos^2\theta-\omega_g\cos^3\theta) e^{i\Phi}\,, \\ 
    &\Box_{\eta} A^{(1)z} =-E_{0y}h_{0+}\omega_g(\sin\theta\cos\theta+\sin^3\theta)e^{i\Phi}\,,\\ 
    &\Phi=K_{\mu}x^{\mu}=k_g y\sin\theta  +(k_a+k_g\cos\theta)z -(\omega_a+\omega_g)t\,,
\end{align}
\end{subequations}
where $\Box_{\eta} \equiv {\eta}^{\nu \rho} \partial_{\nu} \partial_{\rho}$ and the term $\Phi$ denotes the total phase of the $A^{(1)\mu}$. Analyzing these four equations, we find that when $\theta=0$ (the GW and EMW propagate in the same direction), the equations reduce to homogeneous equations with no particular solutions. This is consistent with the conclusion that EMWs and GWs do not interact when propagating in the same parallel direction. 

To obtain the analytical solution in infinite space, we employ the standard retarded Green's function,
\begin{equation}
    G(\vec{r},t; \vec{r}',t') =\frac{-\delta(t-t'-|\vec{r}-\vec{r}'|)}{4\pi|\vec{r}-\vec{r}'|}\,.
\end{equation}
We assume the interaction begins at $t=0$ and impose trivial initial conditions for the first-order field, namely $A^{(1)\mu}(\vec{r}, 0)=0$ and $\partial_t A^{(1)\mu}(\vec{r}, 0)=0$. By inserting the effective source into the integral, the integration naturally resolves into a combination of driven and natural oscillatory modes that satisfy the prescribed boundaries. This yields the explicit solution
\begin{equation}
    A^{(1)\mu} =\frac{1}{2K}J^{\prime\mu}_{\text{eff}}\left[\frac{e^{i(\Omega+K)t}-1}{\Omega+K}-\frac{e^{i(\Omega-K)t}-1}{\Omega-K}\right]\,,\label{A1UP}
\end{equation}
where $K=\sqrt{k_g^2+k_a^2+2k_a k_g\cos\theta}$ and $\Omega=\omega_a+\omega_g$, and $J^{\prime\mu}_{\text{eff}}$ denotes the effective source on the right-hand side of Eqs.~(\ref{EQ}).

After obtaining the contravariant components of the first-order four-potential $A^{(1)\mu}$ in Eq.~(\ref{A1UP}), the corresponding covariant components $A^{(1)}_{\mu}$ are derived using the first-order index-lowering relation (Eq. \ref{RLA}). The first-order EM field tensor $F^{(1)}_{\mu\nu}$ is then constructed from the definition in Eq.~(\ref{F-A}). Likewise, the contravariant tensor $F^{(1)\mu\nu}$ is obtained through the first-order index-raising relation (Eq.\ref{Fsj}).  Once the EM field tensor is determined, the explicit components of the first-order electric and magnetic fields, $E^{(1)}_{\mu}$ and $B^{(1)}_{\mu}$, can be obtained from Eqs.~(\ref{E1}) and (\ref{B1}).
\begin{figure}
    \centering
    \includegraphics[width=1\linewidth]{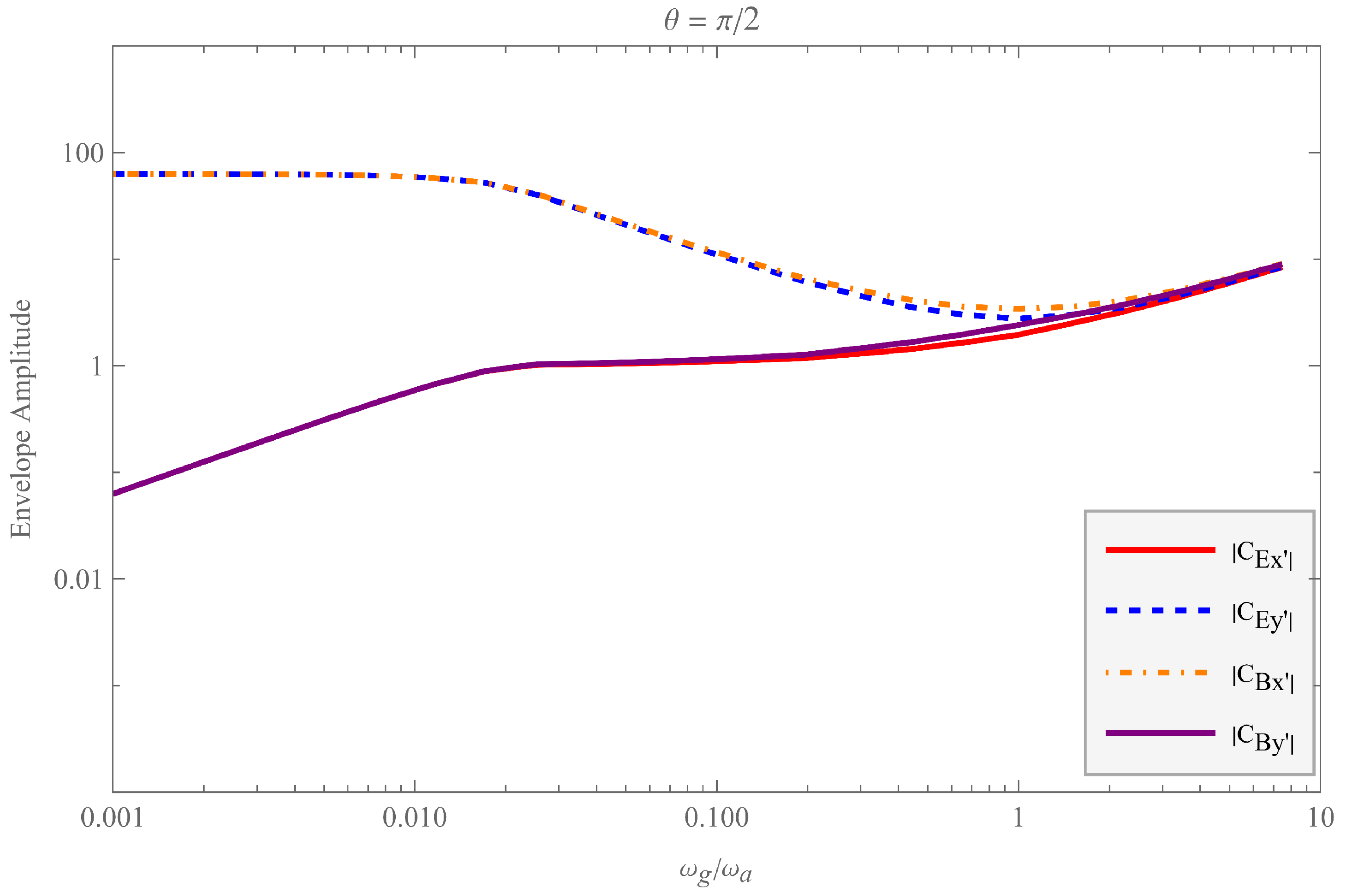}
    \caption{Envelope amplitudes of the field coefficients $|C_{Ex'}|$, $|C_{Ey'}|$, $|C_{Bx'}|$, and $|C_{By'}|$ as a function of the frequency ratio $\omega_g/\omega_a$ for the orthogonal interaction scenario ($\theta = \pi/2$). The vertical axis represents the stable envelope amplitudes extracted over a prolonged interaction time.}
    \label{fig:A4.png}    
\end{figure}
The propagation direction of the coupled mode is characterized by the total wave vector
\begin{equation}
    \mathbf{K}
    = k_g \sin\theta \,\mathbf{e}_y
    + \left(k_a+k_g\cos\theta\right)\mathbf{e}_z \,.
\end{equation}
To align the new $z'$-axis with the direction of $\mathbf{K}$, we apply a rotational transformation. Furthermore, since the total phase $\Phi$ does not satisfy the photon dispersion relation, we hold the wave vector invariant and modify the temporal component of the phase to restore the dispersion relation. By absorbing this temporal correction into the amplitude coefficients, the transformed electric and magnetic fields can be expressed as follows

\begin{subequations}\label{eq_EB_new}
\begin{align}
    &E^{(1)}_{x'}=i C_{Ex'}E_{0y}h_{0\times}e^{i\Phi'}\,,\quad E^{(1)}_{y'}=i C_{Ey'}E_{0y}h_{0+}e^{i\Phi'}\,,\\
    & B^{(1)}_{x'}=i C_{Bx'}E_{0y}h_{0+}e^{i\Phi'}\,,\quad B^{(1)}_{y'}=i C_{By'}E_{0y}h_{0\times}e^{i\Phi'}\,,\\
    &\Phi'= K z' -\Omega' t\,,\\
    & K =\Omega'= \sqrt{\omega_g^2+\omega_a^2+2\omega_a \omega_g\cos\theta}\,.
\end{align}
\end{subequations}
Here, $\Phi'$ represents the modified phase that strictly satisfies the photon dispersion relation, with $\Omega'$ acting as the effective frequency, and the coefficients corresponding to the individual field components, namely $C_{Ex'}, C_{Ey'}, C_{Bx'}, C_{By'}$, exhibit a similar analytical structure. In general, they can be expressed in the form
\begin{equation}
    [\Psi_1
    +\Psi_2e^{i(K-\Omega)t}+\Psi_3e^{i(K+\Omega)t}]e^{i\delta t}\,,
\end{equation}
where the overall factor $e^{i\delta t}$ accounts for the temporal phase correction required to satisfy the dispersion relation, with the frequency shift defined as $\delta=\sqrt{\omega_g^2+\omega_a^2+2\omega_a \omega_g\cos\theta}-\Omega$. The parameters $\Psi_1$, $\Psi_2$, and $\Psi_3$ are functions strictly determined by the EMW frequency $\omega_a$, the GW frequency $\omega_g$, and the interaction angle $\theta$. This analytical form decomposes the dynamic response of the system: the terms within the brackets isolate a time-independent interaction amplitude ($\Psi_1$) from the driven oscillatory modes ($\Psi_2$ and $\Psi_3$), all of which are globally modulated by the temporal phase correction $e^{i\delta t}$. The complete and explicit analytical expressions for these terms are relegated to Appendix~\ref{Appendix A}.

To provide a clearer analysis of the coefficients for the individual field components, Fig.~\ref{fig:A4.png} illustrates the stable envelope amplitudes of their magnitudes over a prolonged interaction time at $\theta = \pi/2$, which corresponds to an orthogonal interaction between the EMW and the GW. As depicted, the coefficients exhibit distinct scaling behaviors across a wide range of frequency ratios. Specifically, in the regime where the initial EMW frequency is relatively small (corresponding to larger values of $\omega_g/\omega_a$), the envelope amplitudes display a continuously increasing trend. This divergent behavior in the low-frequency limit is a clear manifestation of infrared divergence. Conversely, as the frequency ratio decreases, the amplitudes gradually asymptote to stable values. Moreover, for an initially $y$-polarized EM field, the amplitude coefficients associated with the $y$-polarized first-order mode are significantly more pronounced. Quantitatively, based on the peak envelope amplitudes observed in the graph ($C \sim 10^2$), and assuming a typical astrophysical GW with a dimensionless strain of $h_0 \sim 10^{-21}$, the generated first-order EM field is on the order of $10^{-19}$ relative to the incident field amplitude. Finally, utilizing Eqs.~(\ref{T1}) and (\ref{T2}), the first- and second-order energy-momentum tensors can be evaluated. Because both the background and first-order EM fields are purely oscillatory, the first-order energy-momentum tensor strictly vanishes upon time averaging. Consequently, the lengthy explicit expression for the second-order energy-momentum tensor is omitted here for brevity.

It is important to emphasize that this integral solution does not identically satisfy the Lorenz gauge condition, $\partial_\mu A^{(1)\mu} = 0$.  As demonstrated in Sec.~\ref{gaugesection}, the perturbation equations remain gauge invariant under the Lorenz gauge condition. From a mathematical perspective, restricting the interaction interval to $0<t'<t$ effectively introduces a sudden turn-on of the interaction, which is equivalent to multiplying the effective source term by a Heaviside step function $\Theta(t)$. This truncation implies that the modified effective source no longer strictly satisfies the conservation law $\partial_{\mu}J^{\mu}_{\text{eff}}\neq 0$. Due to this non-conservation of the effective source, the resulting vector potential $A^{(1)\mu}$ acquires nonphysical longitudinal and temporal components.

Despite this apparent violation of the Lorenz gauge condition, we retain this particular solution for the subsequent analysis. The essential justification is that our primary concern is the extraction of physically observable, gauge-invariant quantities. In accordance with the gauge invariance encoded in the Ward identity \cite{1950PhRv...78..182W,1957NCim....6..371T}, one can verify that the first-order four-potential satisfies the momentum-space transversality condition
\begin{equation}
    k^{(1)}_\mu A^{(1)\mu} = 0\,,
\end{equation}
where $k^{(1)\mu}$ denotes the wave vector associated with $A^{(1)\mu}$. This condition guarantees that purely longitudinal gauge degrees of freedom do not contribute to physical observables. Consequently, when constructing the first-order EM field tensor $F^{(1)}_{\mu\nu} = \partial_\mu A^{(1)}_\nu - \partial_\nu A^{(1)}_\mu$ the unphysical gauge-dependent contributions are automatically eliminated. Furthermore, consistent with the decoupling mechanism implied by the Ward identity, one could  explicitly discard the longitudinal components of the electric and magnetic fields along the propagation direction, since these components merely represent unphysical artifacts originating from the gauge violation. % \textcolor{red}{Therefore, the subsequent analysis} is restricted entirely to the physically meaningful transverse field components.

\section{CONCLUSION}

In this work, we have developed a systematic perturbative framework for investigating the first-order EM response induced by GWs in curved spacetime. Starting from the covariant Maxwell equations, we performed a consistent perturbative expansion of both the four-potential and the EM field tensor in tensor form, and derived the corresponding first-order perturbation equations. We further demonstrated that the first-order perturbation equations preserve residual gauge invariance under the Lorenz gauge condition. Based on these results, we derived explicit first-order expressions for the electric field, magnetic field, and EM energy-momentum tensor in curved spacetime.

Crucially, we emphasize that metric-induced corrections arising from index manipulation play an essential role in maintaining the internal consistency of the perturbative formalism. Because the metric itself is perturbed, the operations of raising and lowering indices introduce non-trivial discrepancies between the covariant and contravariant expressions of the fields, which must be meticulously tracked during tensor calculations. Furthermore, the validity of the entire perturbative framework hinges on a self-consistent unperturbed state; therefore, it is imperative that the prescribed background EM field strictly satisfies the zeroth-order Maxwell equations before any perturbative calculations can proceed.

As a concrete application, we analytically investigated the interaction between a plane EMW and a GW in the transverse-traceless gauge. Solving the strict initial value problem via the retarded Green's function method yielded explicit analytical expressions for the first-order EM fields. Crucially, our analysis of the coupling coefficients revealed a distinct infrared divergence in the low-frequency limit of the incident EMW, which gradually asymptotes to stable values at smaller frequency ratios. The interaction predominantly preserves the initial polarization state. Quantitatively, a characteristic GW strain of $h_0 \sim 10^{-21}$ scales the generated first-order EM field to approximately $10^{-19}$ relative to the background. Future work will focus on evaluating the GW perturbation problem under the background of a bounded static magnetic field.

\appendix

\section{Explicit Analytical Expressions for the Amplitude Coefficients}\label{Appendix A}
In this appendix, we provide the complete and explicit analytical expressions for the amplitude coefficients $C_{Ex'}$, $C_{Ey'}$, $C_{Bx'}$, and $C_{By'}$ in Eq. \ref{eq_EB_new}. To present these expressions in a compact and physically transparent form, we substitute the total frequency $\Omega = \omega_a + \omega_g$, the norm of the wave vector $K = \sqrt{\omega_a^2 + \omega_g^2 + 2\omega_a\omega_g\cos\theta}$ and the frequency shift defined as $\delta=\sqrt{\omega_g^2+\omega_a^2+2\omega_a \omega_g\cos\theta}-\Omega$. Furthermore, to avoid repetition in the expressions for $C_{Ey'}$ and $C_{Bx'}$, we define an auxiliary angular function $\Lambda(\theta)$ as follows
\begin{equation}
    \begin{aligned}
        \Lambda(\theta) = &\omega_a^2 + \omega_a\omega_g + \omega_g^2 + (\omega_a^2 + \omega_a\omega_g + 2\omega_g^2)\cos\theta\\
        &+ \omega_a\omega_g\cos(2\theta)\,.
    \end{aligned}
\end{equation}
As discussed in the main text, each coefficient is decomposed into a time-independent interaction amplitude ($\Psi_1$) and two driven oscillatory modes ($\Psi_2$ and $\Psi_3$). The unified structure for the coefficients takes the form
\begin{equation}
    \begin{aligned}
        C_{j} = &\left[ \Psi_{1,j} + \Psi_{2,j} e^{-i(K-\Omega)t} + \Psi_{3,j} e^{i(K+\Omega)t} \right] e^{i\delta t}\,, \\
        &(j = Ex', Ey', Bx', By')\,.
    \end{aligned}
\end{equation}
By simplifying the analytical results derived from the perturbation expansion, the specific expressions for the components $\Psi_{1,j}$, $\Psi_{2,j}$, and $\Psi_{3,j}$ are rigorously determined as follows.

\noindent\textbf{1. Expressions for $C_{Ex'}$}
\begin{align}
    \Psi_{1, Ex'} &= -\frac{i\Omega}{2\omega_a}\,, \\
    \Psi_{2, Ex'} &= \frac{i(K+\Omega)}{4\omega_a}(1+2\cos\theta)\,, \\
    \Psi_{3, Ex'} &= -\frac{i(K-\Omega)}{4\omega_a}(1+2\cos\theta)\,.
\end{align}

\noindent\textbf{2. Expressions for $C_{Ey'}$}
\begin{align}
    \Psi_{1, Ey'} &= \frac{i\Omega(\omega_a^2+\omega_g^2+\omega_a\Omega\cos\theta)}{2K\omega_a\omega_g}\,, \\
    \Psi_{2, Ey'} &= -\frac{i(K+\Omega)\Lambda(\theta)}{4K\omega_a\omega_g}\,, \\
    \Psi_{3, Ey'} &= \frac{i(K-\Omega)\Lambda(\theta)}{4K\omega_a\omega_g}\,.
\end{align}

\noindent\textbf{3. Expressions for $C_{Bx'}$}
\begin{align}
    \Psi_{1, Bx'} &= -\frac{i\Omega(\Omega+\omega_a\cos\theta)}{2\omega_a\omega_g}\,, \\
    \Psi_{2, Bx'} &= \frac{i(K+\Omega)\Lambda(\theta)}{4K\omega_a\omega_g}\,, \\
    \Psi_{3, Bx'} &= \frac{i(K-\Omega)\Lambda(\theta)}{4K\omega_a\omega_g}\,.
\end{align}

\noindent\textbf{4. Expressions for $C_{By'}$}
\begin{align}
    \Psi_{1, By'} &= -\frac{i\Omega(\Omega+2\omega_a\cos\theta)}{2K\omega_a}\,, \\
    \Psi_{2, By'} &= \frac{i(K+\Omega)}{4\omega_a}(1+2\cos\theta)\,, \\
    \Psi_{3, By'} &= \frac{i(K-\Omega)}{4\omega_a}(1+2\cos\theta)\,.
\end{align}

\bibliography{apssamp}% Produces the bibliography via BibTeX.

\end{document}